\documentclass[11pt,a4paper]{article}

\usepackage[top=12mm,bottom=12mm,left=30mm,right=30mm,head=12mm,includeheadfoot]{geometry}
\usepackage{titlesec}
\titlespacing*{\section}{0pt}{1.8\baselineskip}{\baselineskip}

\usepackage[utf8]{inputenc}

\usepackage{doi}

\usepackage{amsmath}

\usepackage{hyperref}

\usepackage{lineno}

\usepackage{fancyhdr}
\makeatletter
  \let\ps@plain\ps@fancy
\makeatother

\usepackage{xcolor}
\definecolor{scipostdeepblue}{HTML}{002B49}
\definecolor{scipostphys}{HTML}{0019A2}
\definecolor{scipostpurple}{HTML}{605AAF}
\definecolor{scipostred}{HTML}{A10800}

\usepackage{graphicx}

\usepackage{cite}

\usepackage[width=.90\textwidth]{caption}

\newcommand*\patchAmsMathEnvironmentForLineno[1]{%
\expandafter\let\csname old#1\expandafter\endcsname\csname #1\endcsname
\expandafter\let\csname oldend#1\expandafter\endcsname\csname end#1\endcsname
\renewenvironment{#1}%
{\linenomath\csname old#1\endcsname}%
{\csname oldend#1\endcsname\endlinenomath}}%
\newcommand*\patchBothAmsMathEnvironmentsForLineno[1]{%
\patchAmsMathEnvironmentForLineno{#1}%
\patchAmsMathEnvironmentForLineno{#1*}}%
\AtBeginDocument{%
\patchBothAmsMathEnvironmentsForLineno{equation}%
\patchBothAmsMathEnvironmentsForLineno{align}%
\patchBothAmsMathEnvironmentsForLineno{flalign}%
\patchBothAmsMathEnvironmentsForLineno{alignat}%
\patchBothAmsMathEnvironmentsForLineno{gather}%
\patchBothAmsMathEnvironmentsForLineno{multline}%
}
\relax

\hypersetup{
    colorlinks,
    linkcolor={red!50!black},
    citecolor={blue!50!black},
    urlcolor={blue!80!black}
}

\usepackage[bitstream-charter]{mathdesign}
\usepackage{float}
\usepackage{caption}
\DeclareSymbolFont{usualmathcal}{OMS}{cmsy}{m}{n}
\DeclareSymbolFontAlphabet{\mathcal}{usualmathcal}

\newcommand{\MSb}{{\overline{\rm MS}}}
\newcommand{\MMSb}{{\rm{M}\overline{\rm MS}}}

\begin{document}

\begin{center}{\Large \textbf{
Twist-3 partonic distributions from lattice QCD\\
}}\end{center}

\begin{center}
Shohini Bhattacharya\textsuperscript{1},
Krzysztof Cichy\textsuperscript{2$\star$},
Martha Constantinou\textsuperscript{1},\\
Andreas Metz\textsuperscript{1},
Aurora Scapellato\textsuperscript{1} and
Fernanda Steffens\textsuperscript{3}
\end{center}

\begin{center}
{\bf 1} Department of Physics, Temple University, Philadelphia, PA 19122 - 1801, USA
\\
{\bf 2} Faculty of Physics, Adam Mickiewicz University, \\ ul. Uniwersytetu Pozna\'nskiego 2, 61-614 Pozna\'n, Poland
\\
{\bf 3} Institut f\"ur Strahlen- und Kernphysik, Rheinische Friedrich-Wilhelms-Universit\"at Bonn, Nussallee 14-16, 53115 Bonn
\\
* kcichy@amu.edu.pl
\end{center}

\begin{center}
\today
\end{center}


\definecolor{palegray}{gray}{0.95}
\begin{center}
\colorbox{palegray}{
  \begin{tabular}{rr}
  \begin{minipage}{0.1\textwidth}
    \includegraphics[width=22mm]{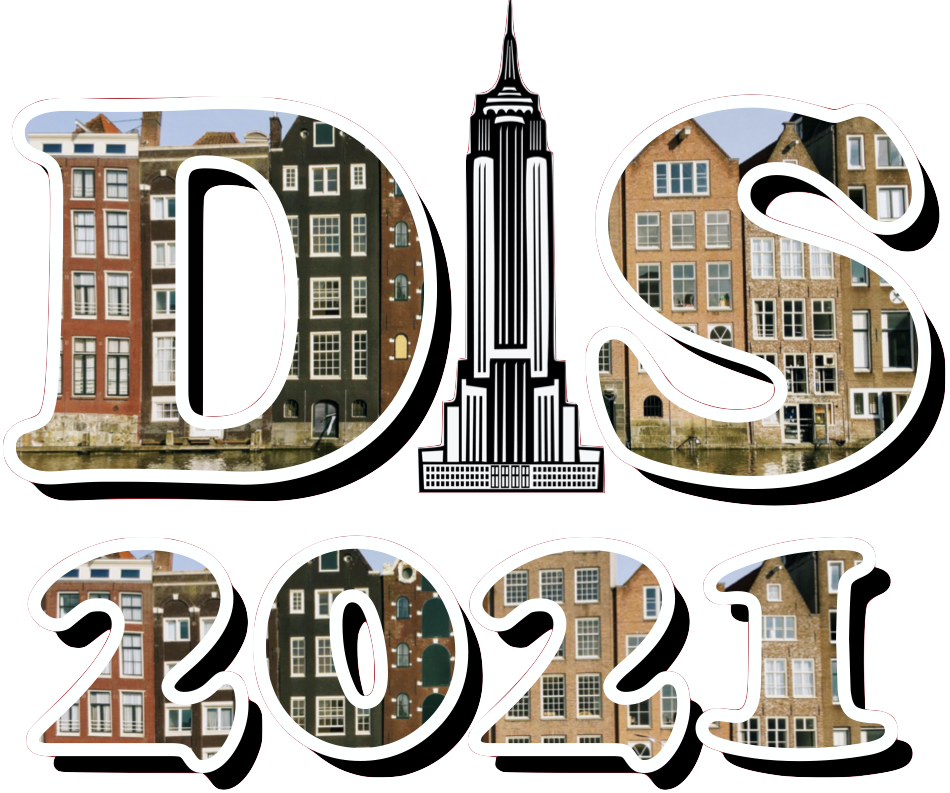}
  \end{minipage}
  &
  \begin{minipage}{0.75\textwidth}
    \begin{center}
    {\it Proceedings for the XXVIII International Workshop\\ on Deep-Inelastic Scattering and
Related Subjects,}\\
    {\it Stony Brook University, New York, USA, 12-16 April 2021} \\
    \doi{10.21468/SciPostPhysProc.?}\\
    \end{center}
  \end{minipage}
\end{tabular}
}
\end{center}

\section*{Abstract}
{\bf
Twist-3 partonic distributions contain important information that characterizes nucleon's structure. In this work, we show our lattice exploration of the twist-3 PDFs $g_T(x)$, and $h_L(x)$. We also present our preliminary results on the twist-3 GPD $\tilde{G}_2(x)$. We use the quasi-distribution approach to connect the lattice-extracted matrix elements, renormalized in the RI/MOM scheme, to light-cone distributions, applying the matching procedure that we developed in parallel. We also calculate the twist-2 counterparts of $g_T(x)$ and $h_L(x)$, i.e.\ $g_1(x)$, and $h_1(x)$, and test the Wandzura-Wilczek approximation.
}


\section{Introduction}
\label{sec:intro}
One of the simplest and yet most important quantities characterizing hadron structure are parton distribution functions (PDFs).
They can be classified according to their twist, i.e.\ mass dimension minus spin of the operator.
The twist determines the order in the inverse scale of the process $Q$ at which they appear in the factorization.
While leading twist (twist-2) distributions are kinematically most important, the higher-twist ones are also vital for full characterization of hadron structure and their relevance is increasing as more precise experimental measurements are becoming available.
Presently, higher-twist distributions are rather poorly known.
While they lack the density interpretation of the twist-2 case, their magnitude can be similar to the latter.
In particular, twist-3 PDFs contain information about quark-gluon-quark correlations and have interesting relations to transverse-momentum-dependent PDFs.
As such, it is important to fill the gaps in our knowledge of higher-twist distributions and we attempt to shed some light on the twist-3 case in our work, by using first-principle lattice QCD simulations.

Lattice QCD is the only known formulation for a systematic \textit{ab initio} study of QCD properties, with the QCD Lagrangian as its starting point and its parameters, i.e.\ quark masses, as the only input.
It is, however, formulated in Euclidean spacetime, which has important consequences for studies of partonic properties.
In fact, for a long time it was considered impossible to directly determine partonic distributions on the lattice, due to their genuine Minkowski spacetime definition.
Lattice computations were restricted only to low moments of such distributions that can be related to Euclidean matrix elements.
Proposals to access the full $x$-dependence were put forward already in the 1990s, but they encountered practical limitations and an actual breakthrough dates back to the seminal proposal of X.~Ji~\cite{Ji:2013dva,Ji:2014gla} a couple of years ago.
In this approach, one calculates spatial (but non-local) analogues of the appropriate light-cone matrix elements that can be Fourier-transformed to $x$-space to define the so-called quasi-distributions.
Quasi-distributions have the crucial property of sharing the infrared physics with their light-cone counterparts and thus, they can be related to the latter (``matching'') by subtracting the difference emerging in the ultraviolet, computable in standard QCD perturbation theory.
Ji's proposal sparked a lot of theoretical and practical interest, as well as revival of some of the early approaches and inception of alternative ones.
For a review of all these efforts, see Refs.~\cite{Cichy:2018mum,Ji:2020ect,Constantinou:2020pek}. 

Until recently, the focus of these investigations was on leading-twist distributions.
The first attempts to address twist-3 PDFs came in 2020 with a series of our three papers \cite{Bhattacharya:2020xlt,Bhattacharya:2020cen,Bhattacharya:2020jfj}.
In Ref.~\cite{Bhattacharya:2020xlt}, we derived the matching for the chiral-even twist-3 PDF, $g_T(x)$, which was the necessary ingredient for its lattice computation in Ref.~\cite{Bhattacharya:2020cen}.
For the other two twist-3 PDFs, the chiral-odd $h_L(x)$ and $e(x)$, matching was derived in Ref.~\cite{Bhattacharya:2020jfj}, taking into account nontrivialities emerging from singular zero-mode contributions.
The $h_L(x)$ function was very recently determined on the lattice \cite{Bhattacharya:2021moj}.
Also recently, an investigation appeared of Burkhardt-Cottingham-type sum rules relating twist-3 PDFs to their twist-2 counterparts in the quark target and Yukawa models \cite{Bhattacharya:2021boh}.
In this paper, we summarize our results from the lattice investigations of $g_T$ and $h_L$.
Additionally, we report on our preliminary investigation of the twist-3 GPD $\tilde{G}_2(x)$ \cite{Kiptily:2002nx} at zero skewness, which is part of the GPD $\tilde{H}(x)+\tilde{G}_2(x)$ that becomes $g_T(x)$ in the limit of zero momentum transfer ($Q^2=0$).
The studies of the matching are reported in another proceedings of this conference \cite{Bhattacharya:DIS21}.

\section{The method and lattice setup}
The twist-3 quasi-distributions can be obtained from the following matrix elements:
\begin{equation}
{\mathcal M}_\Gamma^{\Pi_\mu}(z,P,P')\,=\,\langle P' \,\vert\, \overline{\psi}(0)\,\Gamma W(0,z)\,\psi(z)\,\vert P\rangle\,,
\label{eq:ME}
\end{equation}
where $\vert P\rangle$ is a boosted nucleon state labeled by its 4-momentum, $P=(iE,0,0,P_3)$, $W$ is a Wilson line of length $z$ in the boost direction and we always consider the isovector combination $u-d$. 
We use up to four projectors $\Pi_\mu$, where $\Pi_0=(1+\gamma_0)/4$ and $\Pi_i= (1+\gamma_0)i\gamma_5\gamma_i/4$ ($i=1,2,3$).
The PDF $g_T(x)$ can be obtained from the Dirac structure $\Gamma=\gamma_j\gamma^5$ projected with $\Pi_j$ and we average over the two possible choices ($j=1,2$) to increase statistics.
In turn, its off-forward generalization, $\tilde{G}_2(x)$ at zero skewness, is obtained also from $\Gamma=\gamma_j\gamma^5$ projected with $\Pi_j$ if momentum transfer is perpendicular to both the boost direction and the $j$-direction, and additionally one needs $\Pi_0$ to disentangle the contributions of $\tilde{G}_2$ and the subleading $\tilde{G}_4$.
To access the two remaining GPDs, the two other projectors $\Pi_i$ would be required and here, we report only results for $\tilde{G}_2$, while the other cases will be presented in an upcoming publication.
Finally, $h_L(x)$ is obtained from $\sigma_{12}$ projected with $\Pi_3$.

The bare matrix elements are subject to non-perturbative renormalization in the RI$'$ scheme and converted to a modified $\MSb$ scheme ($\MMSb$) at the scale of 2 GeV.
For GPDs, renormalized matrix elements are additionally subject to disentanglement of $\tilde{G}_2$ and $\tilde{G}_4$.
With our choice of conventions, this results in decomposed matrix elements corresponding to $\tilde{G}_4$ and $\tilde{G}_2+\tilde{H}$, where $\tilde{H}$ is one of the helicity twist-2 GPDs.
Further, we proceed with the reconstruction of the $x$-dependence by employing the Backus-Gilbert method.
For all details of these procedures, we refer to the original publications \cite{Bhattacharya:2020cen,Bhattacharya:2021moj}.
The final stage is to match the ensuing quasi-distributions to their light-cone counterparts in the $\MSb$ scheme at 2 GeV, see Refs.~\cite{Bhattacharya:2020xlt,Bhattacharya:2020jfj,Bhattacharya:DIS21}.

The lattice gauge field configurations used in this work were generated by the ETM collaboration (ETMC)~\cite{Alexandrou:2021gqw} and employ $N_f=2+1+1$ twisted mass fermions with degenerate light quarks leading to a pion mass of 260 MeV and a dynamical strange and charm quark with masses close to the physical ones. The lattice has $32^3\times 64$ sites and a lattice spacing of $a\approx0.093$ fm.
For more details of the lattice techniques, we again refer to Refs.~\cite{Bhattacharya:2020cen,Bhattacharya:2021moj}.

\vspace*{-3mm}
\section{Results}
We start with our results for the twist-3 PDFs, computed at 3 nucleon boosts, $aP_3=4\pi/L,\,6\pi/L,$ $8\pi/L$, corresponding in physical units to 0.83, 1.25 and 1.67 GeV, respectively.
All results shown below are in the $\MSb$ scheme at 2 GeV.

\begin{figure}[h!]
    \begin{center} 
    \includegraphics[scale=0.472]{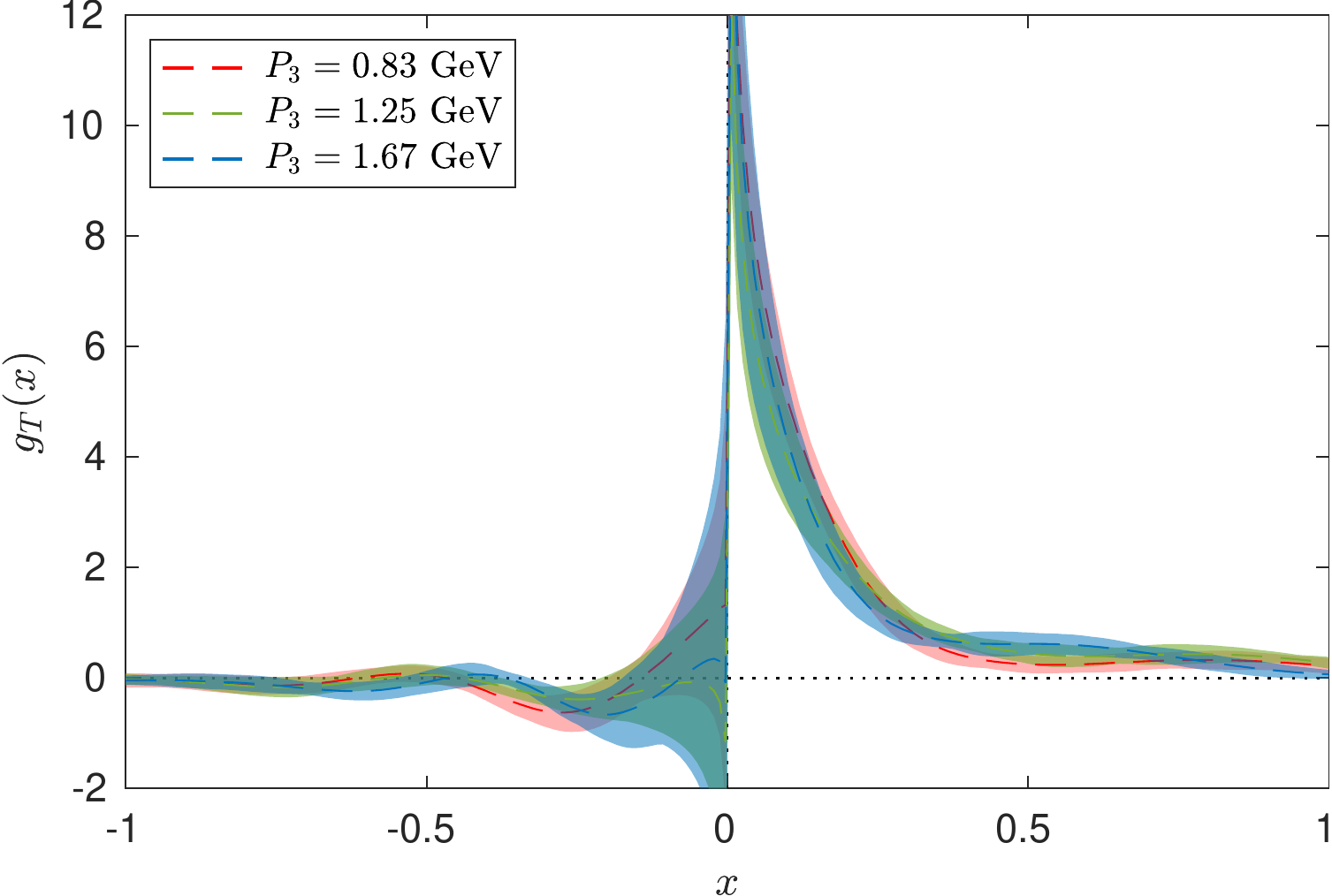}
    \includegraphics[scale=0.472]{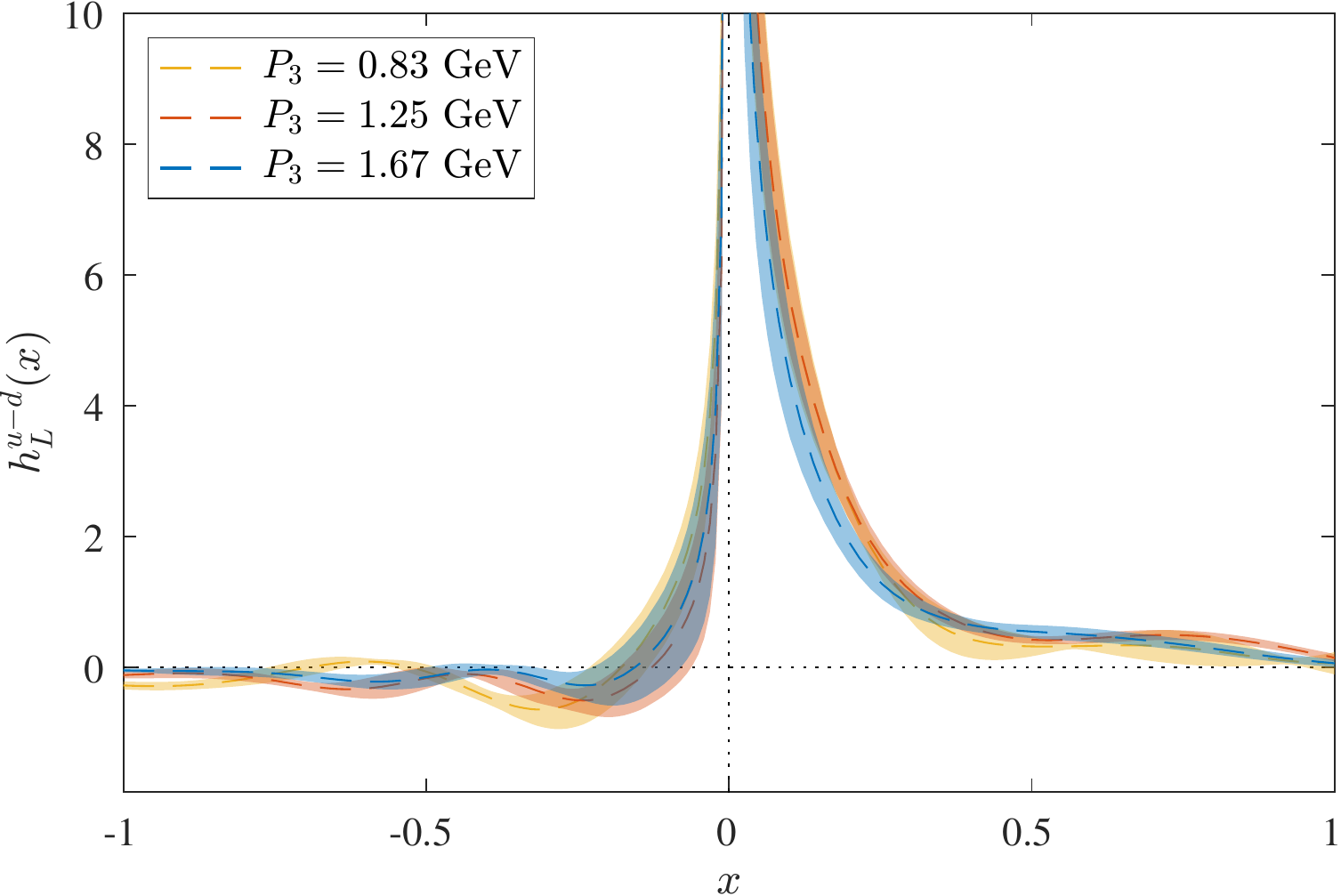}
    \end{center}
    \vspace*{-0.6cm}
    \caption{Nucleon momentum dependence of $g_T(x)$ (left) and $h_L(x)$ (right).}
    \label{fig:momdep}
\end{figure}

In Fig.~\ref{fig:momdep}, we show the final distributions for all momenta, concluding convergence in $P_3$.
Small differences are visible only between the lower 2 momenta in certain ranges of $x$.

\begin{figure}[h!]
    \begin{center} 
    \includegraphics[scale=0.472]{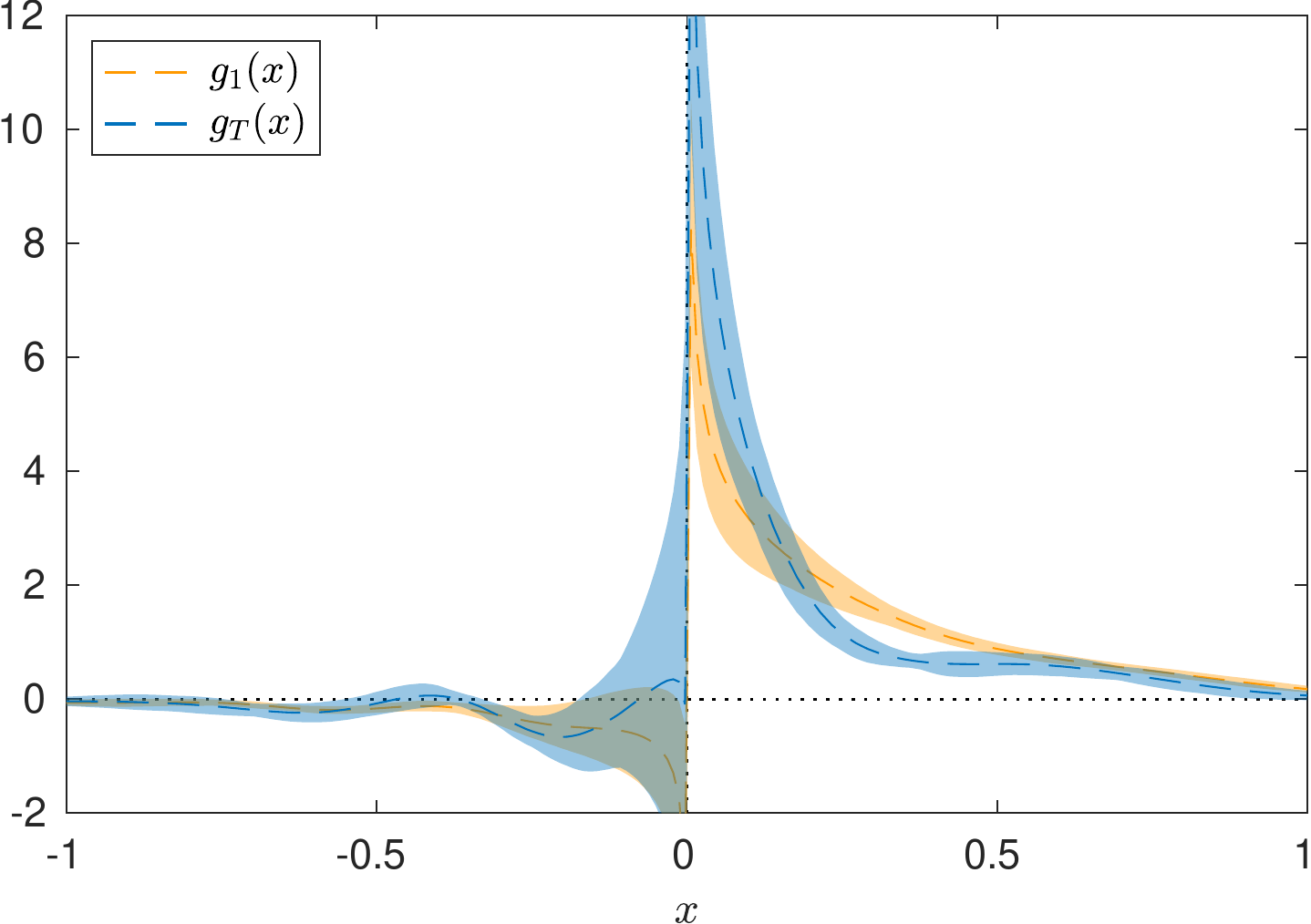}
    \hspace*{3mm}
    \includegraphics[scale=0.472]{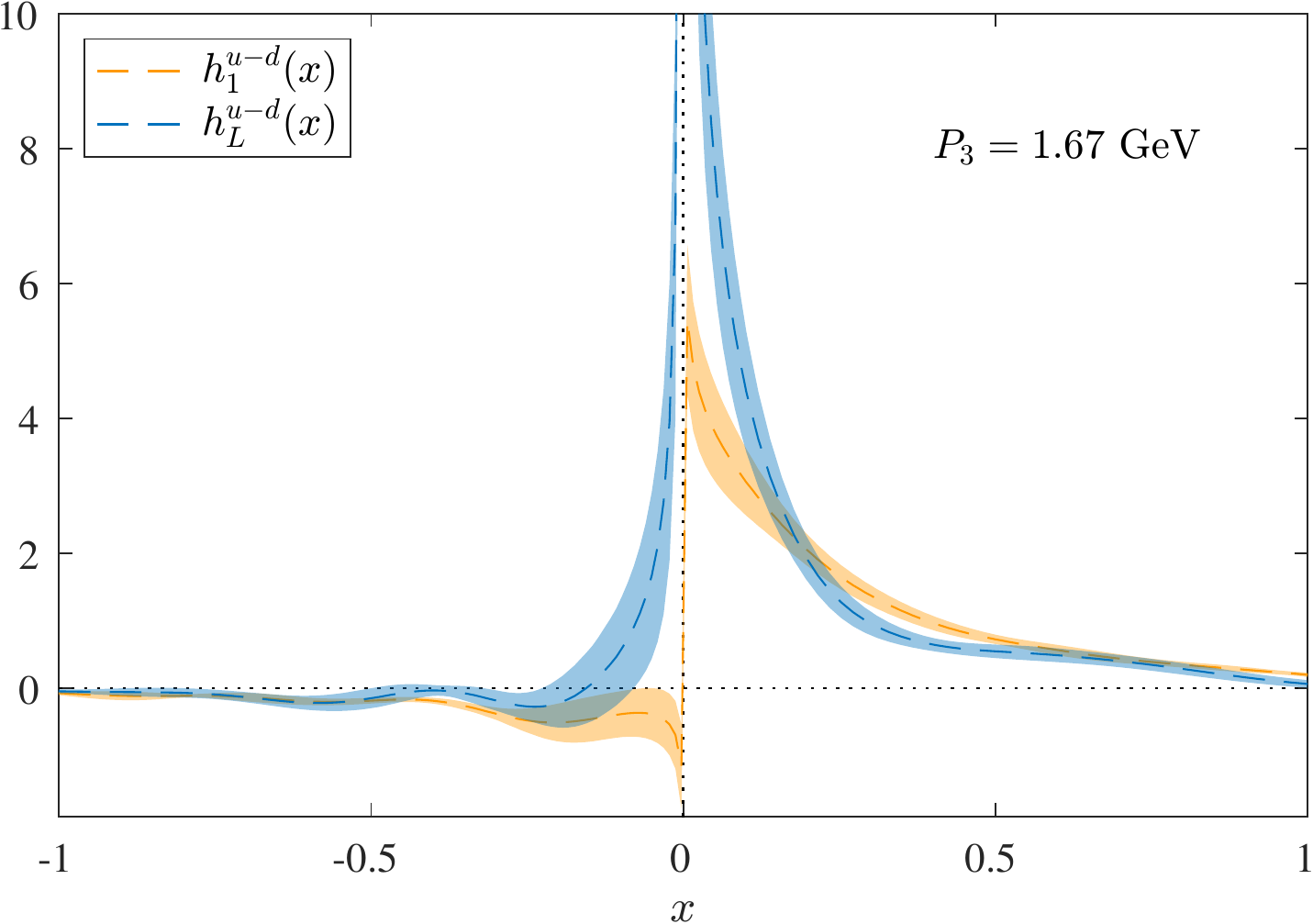}
    \end{center}
    \vspace*{-0.6cm}
    \caption{Comparison of the twist-3 PDFs with their twist-2 counterparts: $g_T(x)$ vs.\ $g_1(x)$ (left) and $h_L(x)$ vs.\ $h_1(x)$ (right).}
    \label{fig:twist32}
\end{figure}

Fig.~\ref{fig:twist32} presents the comparison of the twist-3 PDFs (at our largest boost) with their twist-2 counterparts, the helicity $g_1(x)$ for $g_T(x)$ and the transversity $h_1(x)$ for $h_L(x)$.
In both cases, the twist-3 functions are of similar magnitude, with a tendency for larger values than for the twist-2 PDFs at small-$x$ and a steeper descent in this region.
However, one needs to remember that the region of $x\lesssim0.1$ suffers from enhanced power corrections and hence may be unreliable with the present nucleon boosts and lattice spacings.
We also mention that the Burkhardt-Cottingham sum rules, stating that the twist-3 and the corresponding twist-2 PDFs have the same integrals (equal to the nucleon axial/tensor charge), are satisfied in our data.

\begin{figure}[h!]
    \begin{center} 
    \includegraphics[scale=0.44]{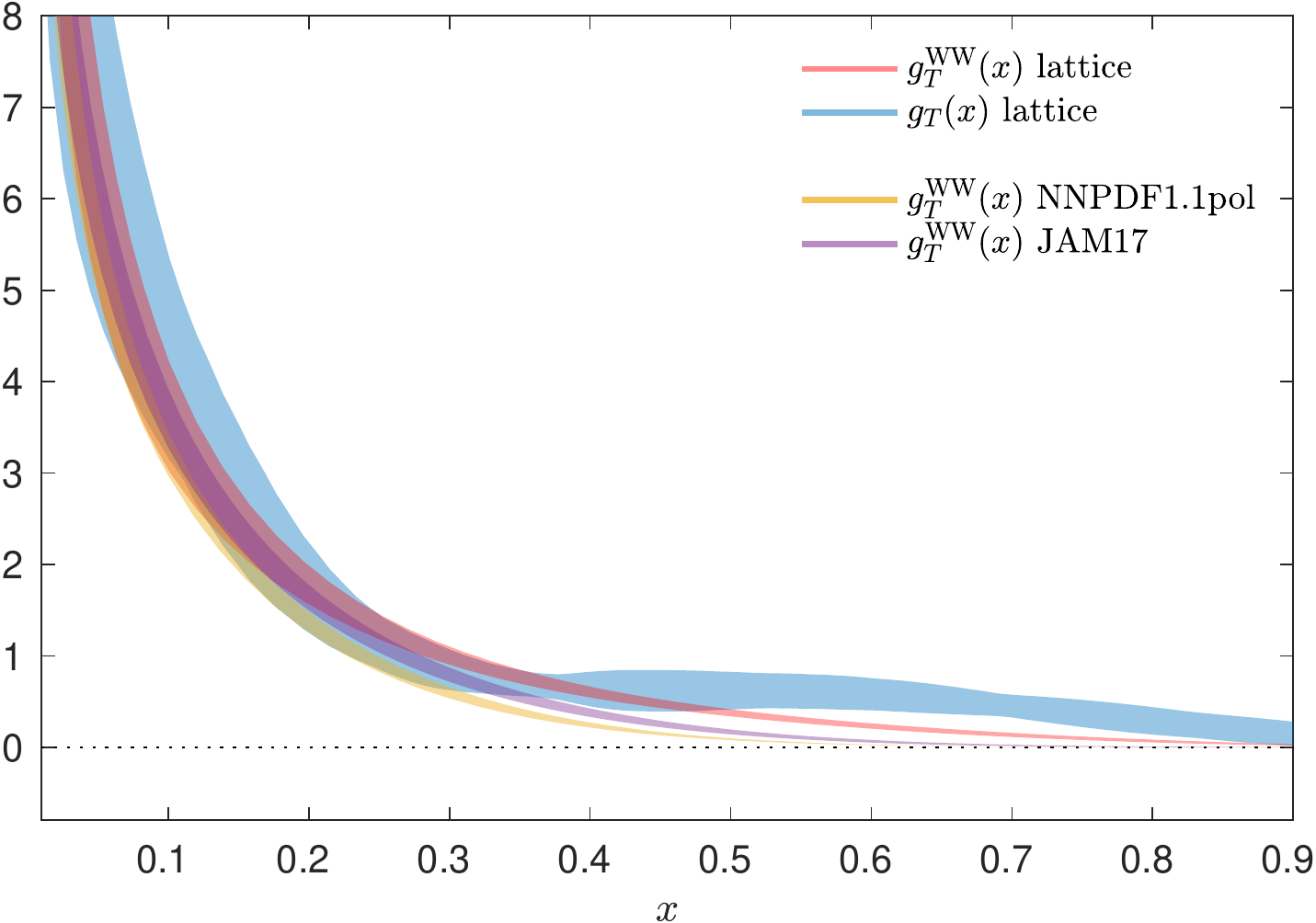}
    \hspace*{3mm}
    \includegraphics[scale=0.44]{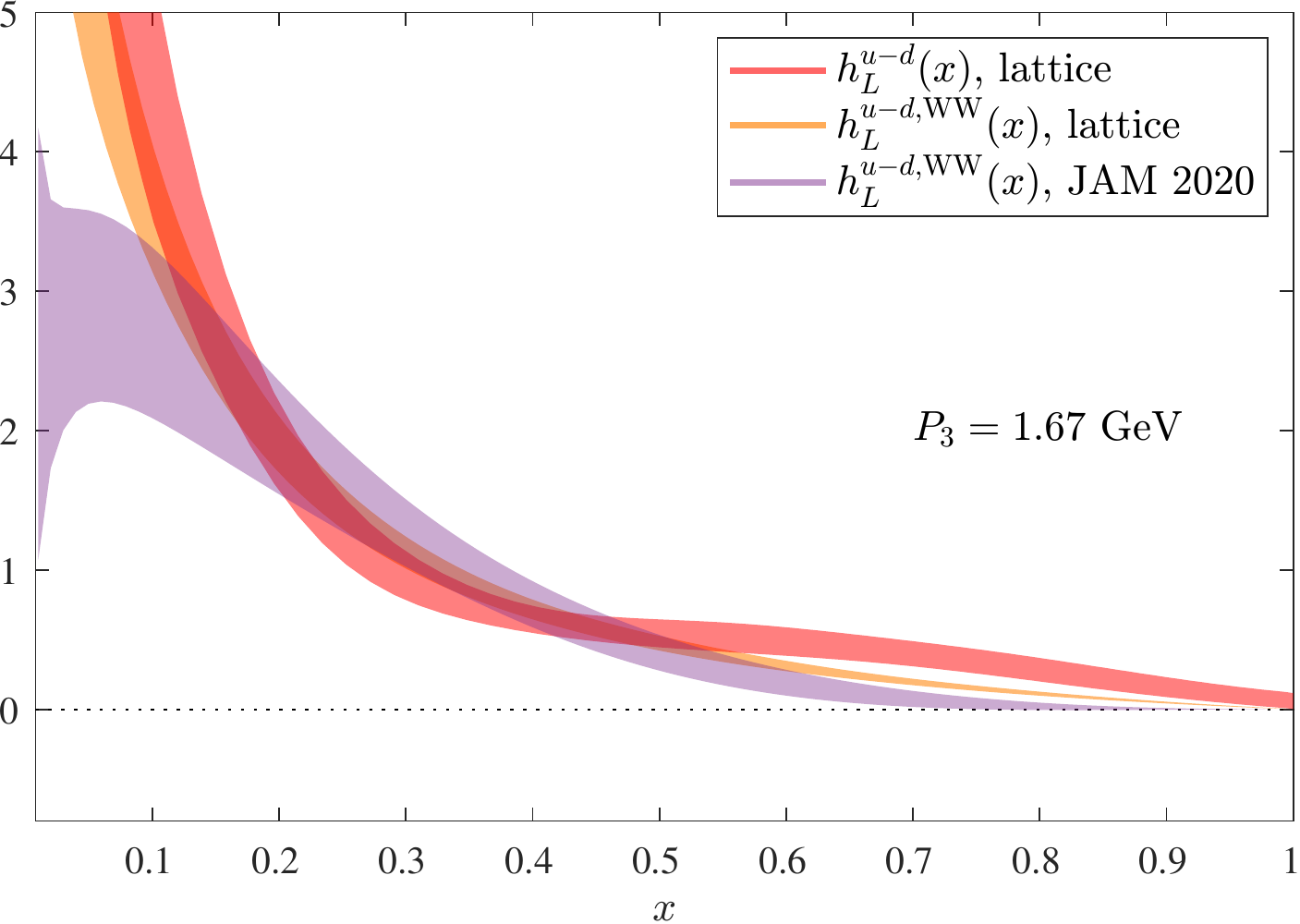}
    \end{center}
    \vspace*{-0.6cm}
    \caption{Test of the WW approximation for $g_T(x)$ (left) and $h_L(x)$ (right).}
    \label{fig:WW}
\end{figure}

An important question in the literature is the validity of the so-called Wandzura-Wilczek (WW) approximation \cite{Wandzura:1977qf, Jaffe:1991ra}.
In this approximation, the twist-3 distributions are fully determined by their twist-2 counterparts, i.e.\ $g_T(x)\approx g_T^{\rm WW}(x)\equiv\int_x^1 g_1(y)\,dy/y$ (where the index WW denotes the function in the WW approximation) and $h_L(x)\approx h_L^{\rm WW}(x)\equiv2x\int_x^1 h_1(y)\,dy/y^2$.
In Fig.~\ref{fig:WW}, we test the WW approximation for both cases, using the lattice-determined $g_1$ and $h_1$ PDFs, and we compare with the WW-approximated functions using data from NNPDF/JAM global fits \cite{Nocera:2014gqa,Ethier:2017zbq,Cammarota:2020qcw}.
We observe that the WW approximation is satisfied within uncertainties for $x\lesssim0.5$ in both cases, with statistically significant deviations in the large-$x$ region.
However, it is important to emphasize that the lattice distributions are subject to yet unquantified systematics that need to be scrutinized before concluding.
Nevertheless, it is interesting to note that within the current uncertainties, the possible violations of the WW approximation are similar to ones found from global analyses \cite{Accardi:2009au}, i.e.\ up to around 40\%.\vspace*{1mm}

\begin{minipage}{\linewidth}
\hspace{-0.046\linewidth}
\begin{minipage}{0.55\linewidth}
\hspace*{13pt} Finally, we show our preliminary results on the twist-3 GPD $\tilde{G}_2(x)$ \cite{Kiptily:2002nx} whose combination with the helicity twist-2 GPD $\tilde{H}(x)$ is the off-forward generalization of $g_T(x)$.
We considered one value of the momentum transfer, $Q^2=0.69$ GeV$^2$.
In Fig.~\ref{fig:GPDs}, we show the comparison of the matched GPD $\tilde{G}_2(x)+\tilde{H}(x)$, $\tilde{H}(x)$ (extracted in Ref.~\cite{Alexandrou:2020zbe}) with the twist-3 PDF $g_T(x)$, all at $P_3=1.25$ GeV.
Comparing twist-3 and twist-2 GPDs, we observe similar behavior as for $g_T$ vs.\ $g_1$ PDFs -- $\tilde{G}_2(x)+\tilde{H}(x)$ is enhanced for small-$x$ with respect to $\tilde{H}(x)$ and falls off more steeply.
Similarly, the comparison of $\tilde{G}_2(x)+\tilde{H}(x)$ with its forward limit, $g_T(x)$, reveals the expected behavior -- the former is suppressed by the momentum transfer, as found also for twist-2 GPDs \cite{Alexandrou:2020zbe}.
\end{minipage}
\hspace{0.03\linewidth}
\begin{minipage}{0.4\linewidth}
  \begin{figure}[H]
     \hspace*{-5mm}
     \captionsetup{width=0.98\linewidth}
     \includegraphics[scale=0.44]{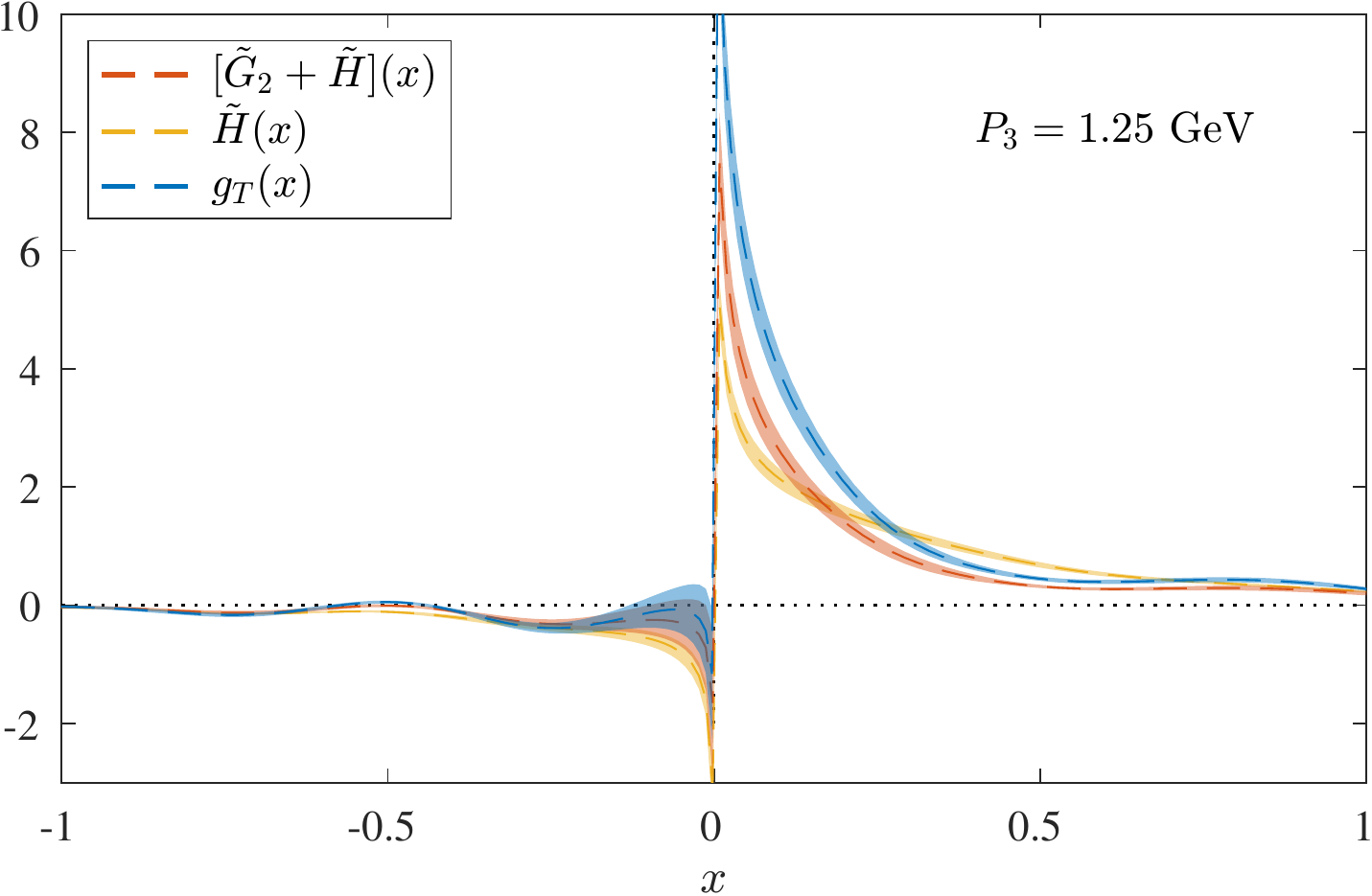}\hspace*{2cm}
     \caption{The matched GPD $\tilde{G}_2(x)+\tilde{H}(x)$ compared to the purely twist-2 helicity GPD $\tilde{H}(x)$ and the twist-3 PDF $g_T(x)$.}
     \label{fig:GPDs}
   \end{figure}
\end{minipage}
\vspace*{0.4mm}
\end{minipage}

\section{Conclusion}
Lattice computations of the $x$-dependence of partonic distributions have seen tremendous progress in the last years.
In addition to the maturing calculations of the simplest of these, twist-2 PDFs, new directions are constantly explored.
In these proceedings, we reported on our exploratory studies of twist-3 PDFs and GPDs.
We showed feasibility of their extractions, but we emphasize that the results are only qualitative as of now.
In the coming years, several sources of systematics need to be explored, both of lattice (e.g.\ finite lattice spacing) and theoretical origin (e.g.\ mixing with quark-gluon-quark operators, see Ref.~\cite{Braun:2021aon}, ignored in our exploratory study).
However, prospects of this direction are clearly encouraging and important first-principle insights into hadron structure are foreseen.

\section*{Acknowledgements}
\begin{footnotesize}
We thank D.~Pitonyak for providing us with the results for the transversity distribution~\cite{Cammarota:2020qcw}.
Computations were carried out in part on facilities of the USQCD Collaboration, which are funded by the Office of Science of the U.S. Department of Energy. 
This research was supported in part by PLGrid Infrastructure (Prometheus system at AGH Cyfronet in Cracow).
Computations were also partially performed at the Poznan Supercomputing and Networking Center (Eagle), the Interdisciplinary Centre for Mathematical and Computational Modelling of the Warsaw University (Okeanos), and at the Academic Computer Centre in Gda\'nsk (Tryton). The gauge configurations have been generated by the Extended Twisted Mass Collaboration on the KNL (A2) Partition of Marconi at CINECA, through the Prace project Pra13\_3304 ``SIMPHYS".
Inversions were performed using the DD-$\alpha$AMG solver~\cite{Frommer:2013fsa} with twisted mass support~\cite{Alexandrou:2016izb}.
\end{footnotesize}

\paragraph{Funding information}
\begin{footnotesize}
The work of S.B.~and A.M.~is supported by the National Science Foundation under grant number PHY-1812359.  A.M.~is also supported by the U.S. Department of Energy, Office of Science, Office of Nuclear Physics, within the framework of the TMD Topical Collaboration. K.C.\ is supported by the National Science Centre (Poland) grant SONATA BIS no.\ 2016/22/E/ST2/00013. M.C.~and A.S.~acknowledge financial support by the U.S. Department of Energy, Office of Nuclear Physics, Early Career Award under Grant No.\ DE-SC0020405. F.S.\ was funded by by the NSFC and the Deutsche Forschungsgemeinschaft (DFG, German Research
Foundation) through the funds provided to the Sino-German Collaborative Research Center TRR110 “Symmetries and the Emergence of Structure in QCD” (NSFC Grant No. 12070131001, DFG Project-ID 196253076 - TRR 110).
\end{footnotesize}

\bibliography{references.bib}

\nolinenumbers

\end{document}